\documentclass{article}

 \usepackage[preprint]{neurips_2025}

\usepackage[utf8]{inputenc} 
\usepackage[T1]{fontenc}    
\usepackage{hyperref}       
\usepackage{url}            
\usepackage{booktabs}       
\usepackage{amsfonts}       
\usepackage{nicefrac}       
\usepackage{microtype}      
\usepackage{xcolor}         
\usepackage{graphicx}
 
\title{Episodic Memory in Agentic Frameworks: Suggesting  Next Tasks}

\author{ 
Sandro Rama Fiorini  \\
IBM Research, Brazil \\
\texttt{srfiorini@ibm.com} \\
\And 
Leonardo G. Azevedo \\
IBM Research, Brazil \\
\texttt{lga@br.ibm.com} \\
\And 
Raphael M. Thiago \\
IBM Research, Brazil \\
\texttt{raphaelt@br.ibm.com} \\
\And 
Valesca M. de Sousa \\
IBM Research, Brazil  \\
\texttt{valesca.sousa@ibm.com} \\
\And 
Anton B. Labate \\
IBM Research, Brazil \\
\texttt{anton.labate1@ibm.com} \\
\And 
Viviane Torres da Silva \\
IBM Research, Brazil \\
\texttt{vivianet@br.ibm.com} \\
}

\begin{document}

\maketitle 

\begin{abstract}
Agentic frameworks powered by Large Language Models (LLMs) can be useful tools in scientific workflows by enabling human-AI co-creation. A key challenge is recommending the next steps during workflow creation without relying solely on LLMs, which risk hallucination and require fine-tuning with scarce proprietary data. We propose an episodic memory architecture that stores and retrieves past workflows to guide agents in suggesting plausible next tasks. By matching current workflows with historical sequences, agents can recommend steps based on prior patterns. 
\end{abstract}


\section{Introduction}

Recent advances in generative models and autonomous agents have opened new avenues for automating complex scientific workflows. Agentic frameworks help leveraging the reasoning and planning capabilities of Large Language Models (LLMs) to explore design spaces, optimize experimental protocols, and accelerate discovery. Those frameworks provide human-AI co-creation capabilities in which humans and AI agents interact to each other to construct and run scientific workflows.

One of the main research areas in workflows construction is the task recommendation. Although it has been addressed before in multiple works (i.e., \cite{goldstein2021augmenting, sola2023activity}), it has not been applied yet in the context of agentic frameworks. While LLMs can be asked what users can do next in a given context, those suggestions are often based on pre-trained memories that are unspecific to a given domain. In order to ameliorate this problem and minimize the likelihood of hallucination~\cite{GuEtAl:2024:AgentSmithSingle}, it would be necessary to fine-tune the LLM with up-to-date dataset of scientific, proprietary workflows, which is not easy to achieve due to limited availability and cost.

One promising direction is to endow the agents themselves with the capacity to store, recall and communicate past users' interactions compiled as formalized workflows. In this paper we propose the use of \emph{episodic memory} to implement this. Inspired by Cognitive Science, it enables agents to store and retrieve experiences contextualized by the user's intentions, facilitating decision-making  \cite{ramos2025review, sumers2023cognitive}. We use episodic memory to store instances of executed scientific workflows, i.e., a  sequence of executed tasks. Agents can then analyze these workflow memories in contrast with the workflow being constructed by an user in order to suggest the next tasks he or she could try next. 

We introduce an episodic memory architecture designed to support workflow completion within agentic frameworks for Materials Science. Our proposal endows existing agentic architectures with the capability of storing and remembering executed workflows. The architecture features a dedicated agent that leverages specialized tools to compile, record and retrieve workflows using a provenance database. Retrieval is based on a workflow similarity algorithm that uses text embeddings and tool comparison. To recommend subsequent steps, we employ a large language model (LLM) that analyzes the retrieved workflows in relation to the current one. Finally, we demonstrate the utility of our architecture through a collection of illustrative examples in material classification

\section{Architecture and methodology}

\begin{figure*}
    \centering
    \includegraphics[width=1\linewidth]{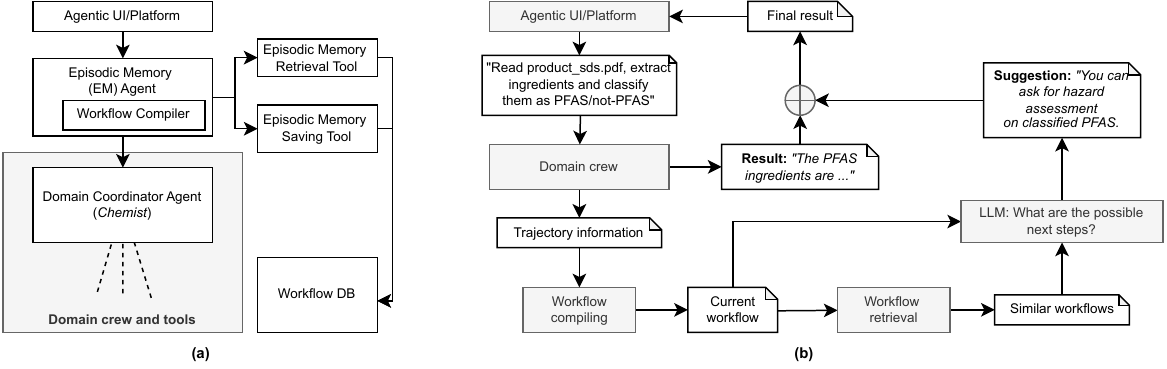}
    \caption{Overview of the approach: (a) general architecture of the solution; (b) flow of the solution, starting on the UI with the user instruction.}
    \label{fig:arch-method}
\end{figure*}

Our task suggestion system is based on the assumption that a user interacts with a crew (i.e., group of agents) of domain agents supervised by a coordinator via a chat interface during a chat session. Instructions and tool calls reflect an intended workflow, even if that is hidden in reasoning steps, reactions and other trajectory information. We also assume that the domain crew is mostly transparent; i.e., we have access to the log of interactions among its sub-agents and tools. Based on these assumptions, we devised the architecture depicted in Figure~\ref{fig:arch-method}a.

The general gist of the method is that we interpose our Episodic Memory (EM) Agent between the chat UI and the domain crew which adds episodic memory capabilities to the system. User instructions are sent to the domain crew, which executes one or more tasks in sequence to fulfill each of the user's instructions. The EM agent compiles a workflow from the trajectory  information coming from the crew (i.e., reasoning steps and tool calls). Based on that information, we retrieve memories of past workflows and infer what could be the next possible steps. 

In more detail, the EM Agent runs a strict task suggestion algorithm (Figure~\ref{fig:arch-method}b). As the user inputs a new instruction in the UI, the EM Agent intercepts and sends it verbatim to the domain agent crew, which provides two outputs: the \emph{task result} itself, which it keeps unchanged, and the \emph{execution trajectory}.

From the trajectory information, the agent uses a \emph{workflow compiler} to create a partial workflow description of the current session (the \emph{current workflow}). In this context, we define a workflow to be a recursive structure composed of ordered steps (i.e., reasoning steps and tool calls), each with input, output and possible nested sub-steps. We distinguish between two types of steps: (a) \emph{function calls}, which are usually tool calls without explicit sub-steps, such as calls to data repositories or simulation functions; and (b) \emph{user-instructions}, which are user-provided instructions that may or may not have tool calls as sub-steps. The process of compiling trajectory into a workflow generates a formalized description of the tasks in the trajectory, ignoring information that is not essential to reproduce the workflow, such as agent reasoning steps and failed tool calls in the domain crew.

After that, the EM Agent runs the \emph{Episodic Memory Retrieval Tool} with the current workflow, which fetches the most relevant workflow memories from a workflow database. The workflow database includes a provenance database with \emph{memory workflows} compiled from interactions with other users, systems or manually curated. We store workflows following the ProvLake provenance schema~\cite{SouzaEtAl:2019:EfficientRuntimeCapture} which is based on W3C PROV~\cite{GrothMoreau:2013:W3CPROVOverview}. The workflow matching algorithm retrieves memory workflows having sub-workflows that are similar to the current one. The agent ignores memory workflows that are exact the same as the current workflow, as they do not provide much information on which steps the user might do next. In our current implementation, we use a simple tree matching algorithm with embedding similarity for instruction steps. More specifically, consider an ordered sequence $S_i$ of the leaf steps of a workflow $W_i$ in the episodic memory. Analogously, let $S_c$ bet the ordered sequence of the leaf steps in the current workflow $W_c$. Considering a similarity threshold $T$, we retrieve $W_i$ if $sim(S_c, S'_i) > T$  for a subsequence $S'_i$ of $S_i$, such that $sim(S_c, S'_i)$ is the mean of the similarities of each step in both sequences. Similarity of user instruction steps is done by cosine similarity of their embeddings; and function call steps are matched if they have the same function name. This scheme gives higher weight to similar sequences of tools, but allows similarity comparison if workflow steps done by agents using no tools.

With the most similar workflows in hand, the agent builds a prompt with textual representations of the selected and current workflows and ask an LLM to produce the most likely next steps for the current one. If no similar workflow can be found, the agent provides the LLM with an alternative prompt with the description of the domain crew and ask the LLM to suggest possible next steps based on the crew's capabilities. In both cases, the suggestions are concatenated back to the result from the domain crew and sent to the UI as the final result. Such concatenation seems like a crude way to build a final answer. However, we found that any use of a final response generator might cause LLMs to start solving the suggested tasks instead of just providing the unaltered result and suggestions.

At any point the user might choose to save the current workflow back in the episodic memory. The EM Agent also implements a simple \texttt{\textbackslash{}save} command, which engages the \emph{Episodic Memory Saving Tool}. This tool compiles and saves the current workflow in the Workflow DB for further use. 

\section{Application}

In order to demonstrate our EM Agent, we constructed a crew for hazard assessment in Material Science. The problem space is identification of PFAS in existing products and assessment of their persistence, bioaccumulation, and toxicity (PBT). Per- and polyfluoroalkyl substances (PFAS) are common chemicals found in all sorts of products, from household items to industrial parts. They have been suffering increased regulatory pressure \cite{abunada2020overview, martineau2024mitigating}, as they have been found to persist for long periods in the environment and organisms. Accessing hazard for PFAS is an important step in processes for control and substitution of these chemicals.

\begin{figure}
    \centering
    \includegraphics[width=0.4\linewidth]{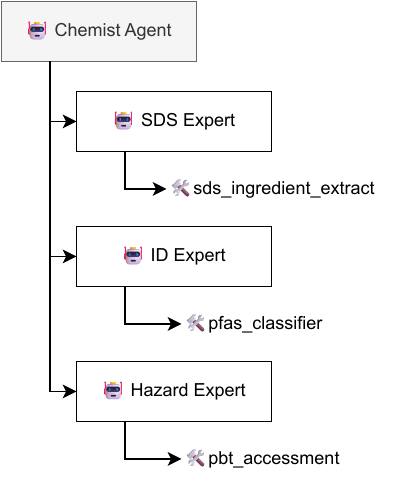}
    \caption{Overview of Chemist Agent and sub-agents in the domain crew. Each sub-agent has a single tool at its disposal and can also infer information by itself.}
    \label{fig:chemist-crew}
\end{figure}

We started by defining a collection of tools for the crew. For that we implemented three tools:
\begin{itemize}
    \item Product extractor: extracts product information from Material Safety Data Sheet (SDS) and Full Material Declaration (FMD) files. These are common files used for describing all kinds of commercialized products and their ingredient chemicals;
    \item PFAS classifier: classifies molecules based on PFAS regulations;
    \item Hazard Assessment: evaluates persistence, bioaccumulation, and toxicity hazard for materials based on \cite{echa2024guidance}.
\end{itemize}
Based on that list, we created a simple LLM based script to bootstrap our episodic memory database. We prompted Llama 4 Maverick FP8\footnote{https://huggingface.co/meta-llama/Llama-4-Maverick-17B-128E-Instruct-FP8} LLM to create a collection of valid possible memory workflows using these tools and two types of simple user instructions.  

\begin{figure}
    \centering
    \fbox{\includegraphics[width=0.6\linewidth]{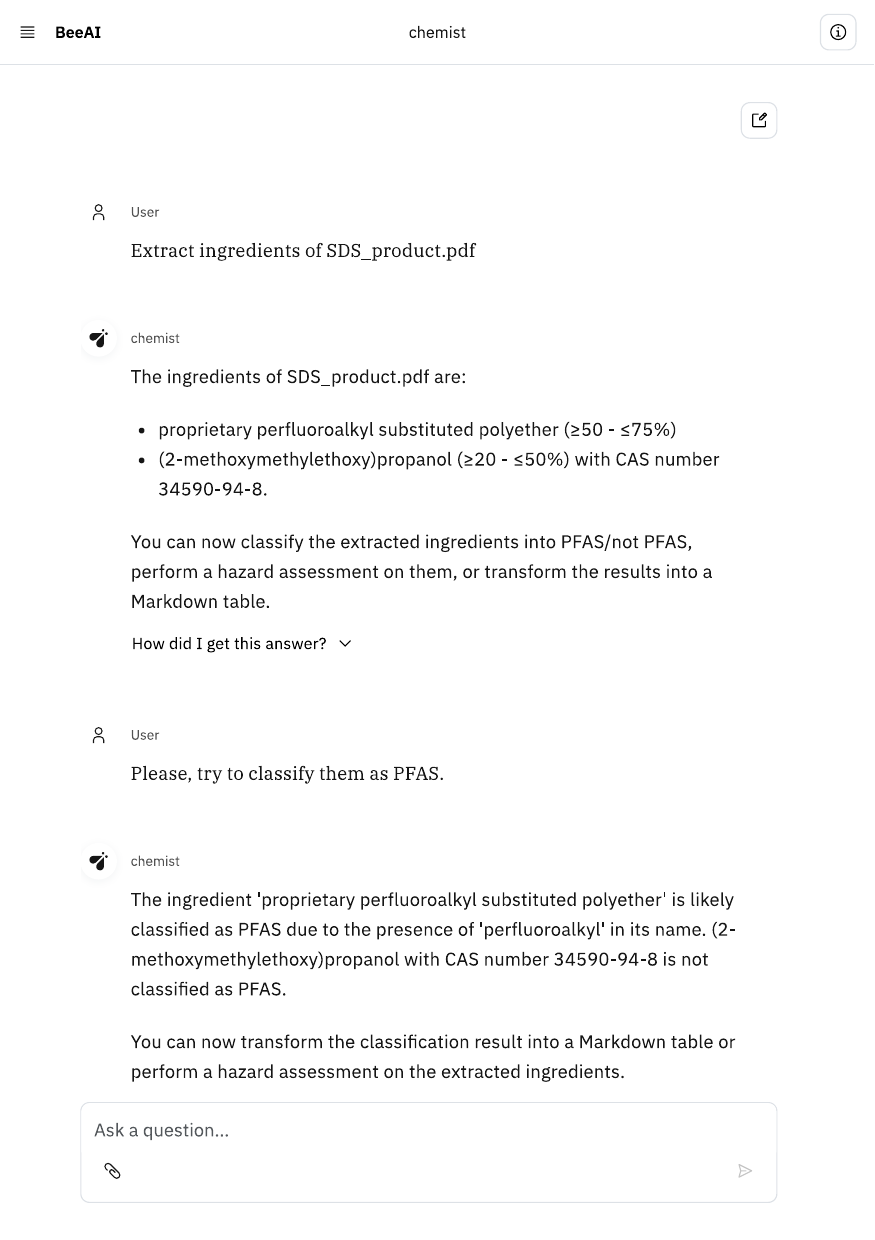}}
    \caption{Chemist crew execution. The chatbot comes up with suggestions based on episodic memory of previous workflows.}        
    \label{fig:beeai-success}
\end{figure}

For testing, we defined an agent crew using the tools above. The architecture can be seen in Figure~\ref{fig:chemist-crew}. We implemented it using BeeAI\footnote{https://github.com/i-am-bee/beeai-framework}. The Chemist Agent coordinates task execution, decomposing the input task into sub-tasks and delegating them to the sub-agent experts. Each of them has a role and a description and a list of tools they may or may not use. The Chemist Agent aggregates the output of its sub-agents and formulate a final text answer to the task. We plugged our EM Agent on top of the Chemist Agent as prescribed in our architecture shown in Figure~\ref{fig:arch-method}a.

\begin{figure}
    \centering
    \fbox{\includegraphics[width=0.6\linewidth]{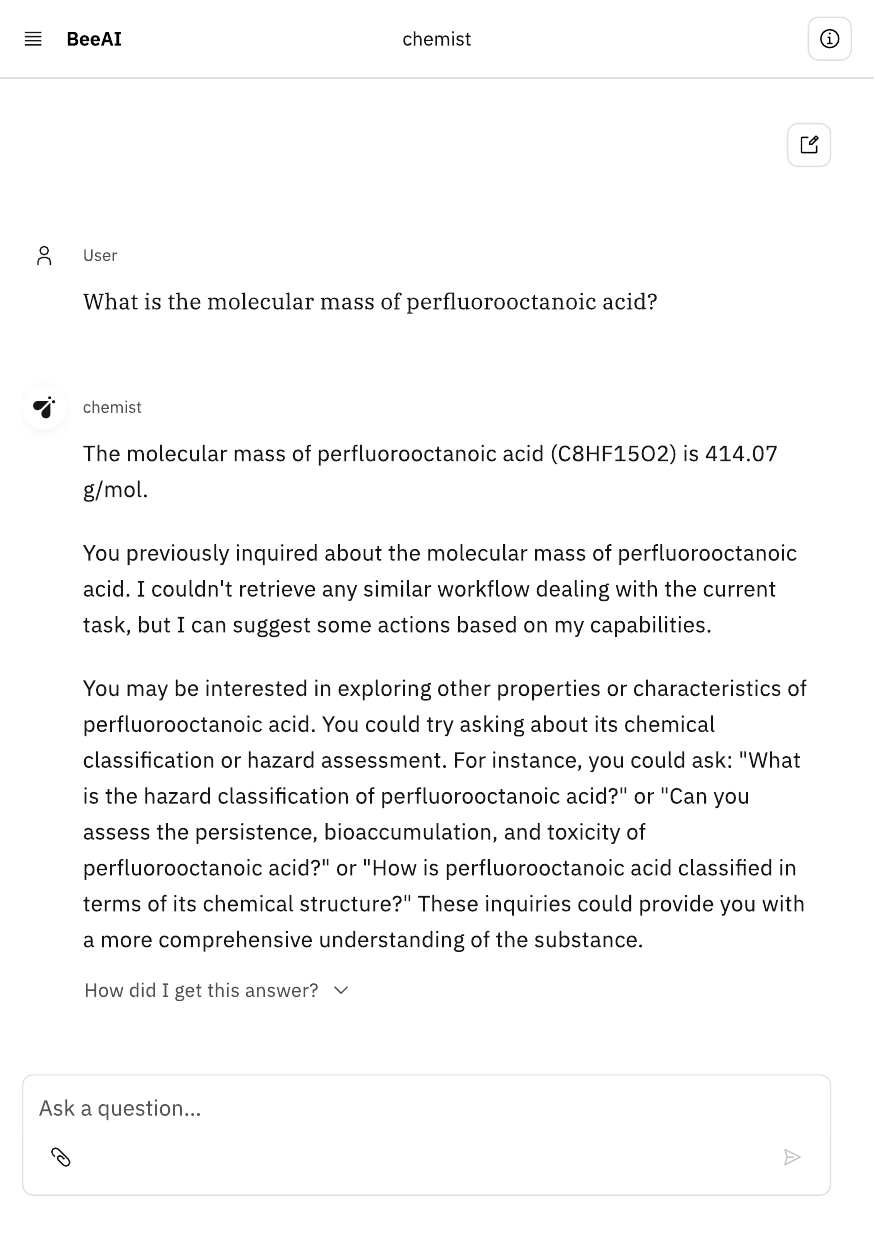}}
    \caption{Execution trace when no suggestion could be made based on episodic memory. The task suggestion follows from the crew description. }
    \label{fig:beeai-failure}
  
\end{figure}

Figure~\ref{fig:beeai-success} shows the output of an incremental execution of the whole system, with two tasks sent to the backend. The user asks it to extract all ingredients of an SDS file --- which describes a product. The crew responds by doing it. Our wrapper agent injects a suggestion of what could be done next. In this case, it finds out that some workflows starting with SDS extraction proceeded to classify ingredients as PFAS, do a hazard assessment or convert the ingredient table to Markdown format. In this implementation, we used $T = 0.65$ as workflow similarity threshold and limited the number of retrieved workflows to a maximum of 10. In this case, the user asks for the PFAS classification. The whole process repeats, taking into account the previous steps as well. Also, note that users can deviate from the suggestions if they wish to do so. The workflow can be saved and re-used for later. Figure~\ref{fig:beeai-failure} shows a case when workflows could not be retrieved, as there is no query or search tool in the crew, therefore no workflows using it. In that case, the Chemist Agent answers the question about perfluorooctanoic acid --- a class of PFAS chemicals --- from learned memory. The EM Agent was not able to recall any workflows with such step. It goes into fall back mode, asking the LLM to suggest a next step based on a description of the crew and its capabilities.

\section{Final remarks}

The work is ongoing and has some obvious limitations. We expect the performance of the tool to be highly dependent on the quality of the curated workflows database and on the definition of what constitutes a relevant workflow. While we expect some resiliency to changes in the agent and tool structure, bigger changes might require more complex workflow matching algorithms. These are avenues for future work, as well as the addition of suggestions for reuse of previous results and planning.

\bibliographystyle{plain}
\bibliography{bibs}

\end{document}